\documentclass[journal]{IEEEtai}

\usepackage[colorlinks,urlcolor=blue,linkcolor=blue,citecolor=blue]{hyperref}

\usepackage{color,array}
\usepackage[table]{xcolor}
\usepackage{graphicx}
\usepackage{amsmath}
\usepackage{cite}
\usepackage{booktabs}

\begin{document}

\title{Fairness in AI-Driven Recruitment: Challenges, Metrics, Methods, and Future Directions}

\author{Dena~F.~Mujtaba and Nihar~R.~Mahapatra}

\markboth{}
{}

\maketitle

\begin{abstract}
The recruitment process significantly impacts an organization's performance, productivity, and culture. Traditionally, human resource experts and industrial-organizational psychologists have developed systematic hiring methods, including job advertising, candidate skill assessments, and structured interviews to ensure candidate-organization fit. Recently, recruitment practices have shifted dramatically toward artificial intelligence (AI)-based methods, driven by the need to efficiently manage large applicant pools. However, reliance on AI raises concerns about the amplification and propagation of human biases embedded within hiring algorithms, as empirically demonstrated by biases in candidate ranking systems and automated interview assessments. 
Consequently, algorithmic fairness has emerged as a critical consideration in AI-driven recruitment, aimed at rigorously addressing and mitigating these biases. This paper systematically reviews biases identified in AI-driven recruitment systems, categorizes fairness metrics and bias mitigation techniques, and highlights auditing approaches used in practice. We emphasize critical gaps and current limitations, proposing future directions to guide researchers and practitioners toward more equitable AI recruitment practices, promoting fair candidate treatment and enhancing organizational outcomes.
\end{abstract}

\begin{IEEEImpStatement}
Artificial intelligence is increasingly used in recruitment for tasks such as preparing job advertisements, screening resumes, and interviewing candidates. These AI tools improve organizational efficiency and accuracy and reduce human subjectivity; however, unintended algorithmic biases can systematically disadvantage qualified applicants, reinforcing harmful stereotypes, undermining fairness, and reducing transparency in hiring decisions. This paper systematically reviews these biases, examines established methods to detect and mitigate unfairness, and highlights practical auditing approaches. By raising awareness of AI recruitment biases and how they can be addressed, this research supports organizations in improving candidate selection, enhancing employee satisfaction and retention, boosting productivity and economic performance, and strengthening public trust in AI-driven decision-making processes.
\end{IEEEImpStatement}

\begin{IEEEkeywords}
Algorithmic bias, algorithmic fairness, artificial intelligence, ethical hiring, human resource management, recruitment
\end{IEEEkeywords}

\section{Introduction}\label{sec:introduction}
\IEEEPARstart{T}{he} recruitment process is crucial for organizational success, influencing employee selection, organizational culture, and overall productivity \cite{huselid1995impact, becker1996impact}. Traditionally, hiring practices—such as job advertisements, skill assessments, and personality tests—have been extensively studied by human resource (HR) professionals and industrial-organizational (I-O) psychologists \cite{mujtaba2019ethical}. Recently, the recruitment landscape has undergone a significant shift toward digital methods, evolving through three phases: from online applications and digital resumes in the 1990s (\textit{digital recruiting 1.0}), to centralized job aggregation platforms in the 2000s (\textit{digital recruiting 2.0}), and currently, to extensive integration of artificial intelligence (AI), termed \textit{digital recruiting 3.0} \cite{black2020ai}. Today, AI-driven tools are increasingly adopted for diverse recruitment tasks including job advertisement creation, candidate screening, video interviewing, and automated candidate assessments \cite{woods2020personnel, mujtaba2019ethical, black2020ai}.

The increasing reliance on AI in recruitment is driven by its demonstrated effectiveness in various HR tasks. A notable application area is job applicant screening, where digital methods can expedite the hiring process and potentially reduce human biases \cite{woods2020personnel}. Large organizations handling substantial numbers of applications increasingly find automated recruitment methods essential for operational efficiency \cite{black2020ai,google-applicants,woods2020personnel}. AI has also begun to play a significant role in crafting job advertisements, with large language models (LLMs), such as ChatGPT, used to draft outlines of necessary skills and qualifications, potentially attracting a broader range of suitable candidates \cite{shrmchatgpt}. Furthermore, organizations have started employing LLMs to generate interview questions and refine communications with job candidates \cite{shrmchatgpt}. Given the rapid shift towards remote work—accelerated by the COVID-19 pandemic—the use of AI for hiring and recruitment decisions is anticipated to further expand \cite{forman_glasser_lech_2020,scanlon2018four}. Recent statistics indicate significant growth, with a 2019 industry survey reporting that 88\% of organizations globally have experimented with AI in recruitment activities; among these, 41\% employed AI-based chatbots for candidate engagement, 44\% utilized AI for identifying candidates through social media and public data, and 43\% leveraged AI for training recommendations \cite{brin_2021}. With expanding access to cloud computing resources, AI integration in recruitment practices is likely to continue growing, underscoring the need to critically assess fairness and transparency of these methods.

\begin{figure*}[h!]
\centering 
\includegraphics[width=\textwidth]{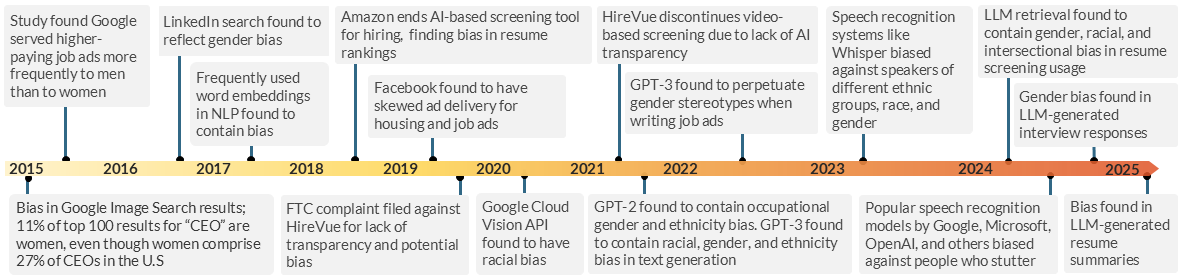}
\caption{Timeline of notable incidents highlighting biases identified in AI systems relevant to or influencing recruitment applications \cite{meyer2018amazon,cooney_2016,miller_2015,caliskan2017semantics,center_2019,ali2019discrimination,feng_has_nodate,bril_2020,hirevuefacial2021,kirk2021bias,abid2021persistent,martin2023bias,fuckner2023uncovering,borchers2022looking,harris2024modeling,kong2024gender,wilson2024gender,seshadri2025does,mujtaba2024lost}.}
\label{fig:timeline}
\vspace*{-.1in}
\end{figure*}

However, as AI adoption becomes more widespread, concerns are growing that decisions made by these systems could be influenced by the biases of organizational personnel or model developers, as evidenced by several recent incidents \cite{mujtaba2019ethical}. For instance, in 2015, Google's job recommendation system exhibited gender bias by displaying high-income job postings more frequently to men than to women \cite{miller_2015}. Similarly, in 2017, Amazon's AI-based candidate evaluation tool was discontinued because it consistently 
assigned lower scores to women's resumes because the AI model had learned gender biases present in historical hiring data, systematically favoring terms common in men's resumes and penalizing those associated with women \cite{mohla2021material,meyer2018amazon}. Moreover, in 2019, Facebook's housing and job ad delivery system faced similar issues, with job ads being skewed based on users' gender and race \cite{ali2019discrimination}. These examples illustrate how biases commonly found in the hiring process, such as hiring discrimination \cite{lewis2018will}, can easily transfer to AI-based systems through the data used for training algorithms. A timeline of well-known instances of bias in AI related to recruitment applications is shown in Figure~\ref{fig:timeline}. Therefore, AI bias in HR applications is already a significant problem, and as AI use continues to grow rapidly, its importance will escalate dramatically in the future.

\textbf{Related Work and Our Contributions:}
\label{sec:related-work}

This paper provides a comprehensive review of fairness in machine learning (ML), deep learning, and AI-based systems specifically for recruitment and hiring, building upon our previous survey \cite{mujtaba2019ethical}. Past studies and surveys have explored biases in non-AI-based recruitment, identifying issues such as \textit{confirmation bias} (making decisions within the first few minutes of meeting a candidate) and \textit{expectation anchor bias} (being influenced by a single piece of information), as well as examining how and where AI can be applied in recruitment \cite{bendick2012developing,van2019marketing,ochmann2019fairness,johnson13Common,laurim2021computer,holm2012recruitment,chen2023ethics}. Additionally, past research has reviewed specific components of the hiring process, such as AI-driven assessments \cite{raghavan2020mitigating}, automated video interviews \cite{hickman2021automated}, and their perceptions by job candidates \cite{chen2023ethics}. Furthermore, recent studies have significantly advanced our understanding of ethical and organizational aspects of AI recruitment, including sociotechnical frameworks for AI adoption in organizations \cite{makarius2020rising}, methods for intelligent talent selection \cite{allal2021intelligent}, challenges of international HR management with AI integration \cite{budhwar2022artificial}, AI capability frameworks in HR management \cite{chowdhury2023unlocking}, and ethical concerns regarding potential dehumanization through AI algorithms in hiring processes \cite{fritts2021ai}. 

However, there remains a need for a systematic review of the literature on algorithmic fairness in AI-based recruitment, encompassing all stages of the hiring process where AI can be applied. This need is underscored by the increasing prevalence of AI in recruitment, such as using LLMs to craft inclusive job descriptions, employing chatbot systems to engage candidates prior to direct recruiter interaction, and integrating AI into applicant tracking systems for resume screening \cite{shrmchatgpt,woods2020personnel,mujtaba2019ethical,black2020ai}. The prevalence is further underscored by a 2022 survey revealing that up to 86\% of employers reported using technology-driven job interviews, with many adopting automated video interviews as a preferred method \cite{hbrAutomatedInterviews}. 

This paper addresses the aforementioned critical gap in the literature by providing a systematic review of algorithmic fairness in AI-based recruitment. It targets AI hiring practitioners and makes several key contributions. First, we explore the definitions of fairness, trust, and justice, drawing from I-O psychology, HR practices, and legislative frameworks to provide a foundational understanding for analyzing bias in AI-driven recruitment (Section~\ref{sec:background}). Second, we discuss the ethical and legal principles governing fairness in AI, highlighting current regulations and guidelines that inform fair AI practices in recruitment (Section~\ref{sec:background:legislation}). Third, we identify and categorize various causes of bias in AI systems, including biases in training data, label definitions, feature selection, proxies, and masking (Section~\ref{sec:causes-of-bias}). Our examination extends across all stages of the recruitment process—sourcing, screening, interviewing, and selection—demonstrating how bias can be introduced and mitigated at each step (Section~\ref{sec:bias}). Fourth, we review diverse fairness metrics, including individual fairness, group fairness, procedural fairness, and fairness in regression, discussing their relevance for employers, jobseekers, and recruitment platforms (Section~\ref{sec:metrics}). Fifth, we categorize and evaluate bias mitigation strategies into pre-processing, in-processing, and post-processing methods, highlighting their significance and application in ensuring fair AI-driven recruitment practices (Section~\ref{sec:methods}). Sixth, we cover the limitations of existing tools and methods for bias detection and mitigation (Section~\ref{sec:detection}), and outline future challenges in this domain (Section~\ref{sec:future}). Finally, Section~\ref{sec:conclusion} presents concluding remarks and summarizes the key findings. By addressing these critical aspects, our paper aims to set a comprehensive foundation for advancing equitable AI practices in recruitment. However, we acknowledge that defining and measuring fairness in recruitment involves complexities due to competing fairness metrics and differing interpretations of employment discrimination law, which will be discussed in subsequent sections.

\textbf{Scope of Review:}
\label{sec:review-scope}
In this survey, we collected relevant articles published by December 2024 through searches on Google Scholar, IEEE Xplore, ACM, ACL, and other publishers. We combined the keywords ``AI'' or ``ML'' with terms such as ``Fairness,'' ``Recruitment,'' ``Recruitment Bias,'' and specific terms representing each recruitment stage: ``Candidate Sourcing,'' ``Candidate Screening,'' ``Candidate Interviews,'' and ``Offer Negotiation.''
The following criteria are used to determine the relevance and/or irrelevance of an article to our survey:
\begin{itemize}
\item Influential or foundational methods or studies in recruitment influencing AI development or understanding fairness in AI. 
\item Work that studies bias, proposes a new method for studying bias, or is applied to mitigate bias in any stage of the recruitment process. 
\item Any external studies and/or related works that are not directly part of recruitment, but can influence AI being used in one of the stages. For example, models such as GPT were not initially created for recruitment, but have heavily influenced the workplace \cite{shrmchatgpt}.
\end{itemize}

\section{Background: Defining Fairness}\label{sec:background}

Before measuring bias, it is essential to define the concept of ``fairness'' to ensure consistent comparison and evaluation of AI for hiring. The following sections provide definitions of fairness within the context of organizational hiring (Section II-A), relevant legislation (Section II-B), and common causes of bias in AI systems (Section II-C).  

\subsection{Relating Fairness to Trust and Justice}\label{sec:background:fairness}
In organizational hiring contexts, fairness cannot be adequately understood without considering its close relationship with trust and justice. A clear conceptual connection among these terms is essential, as misunderstanding this relationship can significantly undermine both the perceived and actual fairness of AI-driven recruitment practices. The terms ``fairness,'' ``trust,'' and ``justice'' are closely related and often used interchangeably \cite{landy2016work}. Before defining fairness, it is useful to examine its relationship with trust and justice, particularly in the context of AI-based hiring.

First, trust, as defined in I-O psychology, is a psychological state characterized by the willingness of an individual to accept vulnerability based on positive expectations regarding the intentions, behavior, and outcomes associated with another entity—whether a person, organization, or technology. It encompasses reliability (consistent performance), integrity (adherence to ethical principles), competence (ability to perform tasks), and benevolence (consideration for others’ welfare) \cite{lee2004trust,dirks2001role,landy2016work,glikson2020human,lockey2021review}. Rather than simply being derived from past interactions, trust also involves contextual factors, perceived risk, regulatory oversight, and individual predispositions towards uncertainty, sometimes described as a ``leap of faith'' \cite{glikson2020human}.
Trust takes time to build and becomes more significant when perceived uncertainty and risk increase (e.g., when the risk of job loss is high) \cite{landy2016work}. However, trust can be easily lost, and once lost, it is very difficult to rebuild, leading workers and employees to question the fairness of organizational actions \cite{landy2016work}. Therefore, trust in AI-based hiring significantly influences a jobseeker’s perception of the fairness of an organization’s recruitment process.

Trust in AI-driven recruitment technology is influenced by the characteristics of the end-user (e.g., the jobseeker's or recruiter's personality and ability), the organizational environment (e.g., institutional policies and workplace culture), and the recruitment system itself (e.g., accuracy, performance, transparency, user reviews, and goal congruence---the alignment between the system's objectives and organizational or individual user goals) \cite{siau2018building}.
Specific factors that impede trust in AI include its role in technological displacement, lack of transparency, biased or unfair decisions, lack of accountability for those responsible for AI decisions, and privacy concerns arising from the collection and reliance on large training datasets \cite{siau2018building,afroogh2024trust,lockey2021review}; for a detailed review of trust in AI, we direct readers to the paper by Afroogh et al. \cite{afroogh2024trust}. A 2019 poll highlights the lack of trust in AI, revealing that 88\% of Americans are skeptical of AI-driven recruitment \cite{shaw2003justify, feffer_2019}, and as of 2023, 71\% opposed AI making final hiring decisions \cite{pewresearch2023}. Another study of recruiters reported they were cautious of the AI's ability to make fair and accurate hiring decisions, and often found job candidates were unwilling to engage with AI assessments, potentially losing highly skilled candidates \cite{li2021algorithmic}. Therefore, even if model developers consider fairness metrics and precautions, a jobseeker’s trust will significantly affect the perceived fairness of any AI-based decisions. It is also important to recognize the inherent ethical considerations and power dynamics within recruitment, as the interests of jobseekers and employers often diverge. An ethical recruitment process should balance these competing interests, rather than solely serving as a mechanism for employers to filter out less-desirable candidates.

Organizational justice, in this context, refers to the just treatment in organizational processes and outcomes, such as recruitment, and is a crucial part to understanding trust in AI because it significantly influences perceptions of fairness \cite{landy2016work}. The literature identifies three primary types of organizational justice: distributive, procedural, and interactional. \textit{Distributive justice} pertains to the perceived fairness of the allocation of outcomes or rewards to individuals within an organization. This can be based on the \textit{merit or equity norm} (i.e., those who work harder or produce more should receive greater rewards), the \textit{need norm} (i.e., rewards are distributed in proportion to the needs of individuals), or the \textit{equality norm} (i.e., all individuals receive equal rewards, regardless of effort or need) \cite{landy2016work,grgic2018beyond}. \textit{Procedural justice} is the perceived fairness of the processes by which rewards are distributed \cite{landy2016work,grgic2018beyond}. Lastly, \textit{interactional justice} concerns the perceived extent to which an organization treats an individual or employee with respect. Interactional justice has two aspects: \textit{interpersonal justice}, which pertains to the extent of respect and politeness accorded to individuals, and \textit{informational justice}, which reflects how well an organization explains its procedures and outcomes to individuals, ensuring that communication is candid, thorough, and timely \cite{landy2016work}. 

In the recruitment process, these aspects of trust and dimensions of organizational justice all factor into how fairness is defined according to AI fairness research and past legislation, as will be further described.

\subsection{Fairness According to Legislation and Ethical Principles}\label{sec:background:legislation}
Considerable progress has been made in establishing standards for fairness and ethical AI. Notably, the FAT/ML (Fairness, Accountability, and Transparency in Machine Learning) community has developed five guiding principles—responsibility, explainability, accuracy, auditability, and fairness—to help developers and product managers design and implement accountable automated decision-making algorithms \cite{diakopoulos2017principles}. The group also recommends adherence to OECD privacy principles and the ethical principles outlined in the Belmont Report for human subjects research to ensure algorithmic accountability \cite{oecd-privacy}.

Additionally, in 2020, the U.S. Department of Defense defined five ethical principles for its AI systems: responsible, equitable, traceable, reliable, and governable \cite{dod_ai_principles,dod_2022_ai}. Following this, numerous companies, including Google \cite{googleai} and OpenAI \cite{openaiSafety}, have established similar guidelines for ethical development and integration of their AI technologies. 

Similarly, legislation regulating technology has emerged to enhance fairness in AI systems. For instance, in 2018, the European Union enforced the General Data Protection Regulation (GDPR) to protect individuals concerning the privacy, processing, and free movement of personal data \cite{gdpr}. Moreover, in 2021, the European Commission established the first legal framework on AI, grouping AI systems into different risk levels \cite{eucai2021,eu_ai_act_2021}; this was officially adopted in 2024 \cite{butt2024analytical}. These levels include minimal risk, such as AI-enabled video games or spam filters that pose minimal or no risk to a citizen's rights or safety; limited risk, where AI systems like chatbots inform users they are interacting with a machine and enable them to make informed decisions before continuing use; high-risk, covering AI systems critical to infrastructure, vocational training, and essential public services like credit scoring, law enforcement, and migration; and unacceptable risk, involving AI systems that present a clear threat to the safety and rights of people \cite{eucai2021}. Similar legislative considerations are being explored internationally; for example, Canada and Australia are currently developing legislation targeting fairness and employee protections in AI-driven recruitment \cite{canada2025ai,dentons2025australia}.

The National Institute of Standards and Technology (NIST) and the Federal Trade Commission (FTC) have also emphasized the need to identify and manage bias in AI \cite{ftc-2021,nistproposalAI2021}. Specifically, the FTC has warned against the sale of racially biased systems that could prevent individuals from obtaining employment, insurance, or other benefits. The FTC highlighted three laws for AI developers and users: Section 5 of the FTC Act, which prohibits unfair or deceptive practices; the Fair Credit Reporting Act, which ensures algorithms do not unfairly deny individuals housing or credit; and the Equal Credit Opportunity Act, which prohibits algorithms from making biased credit decisions based on race, color, religion, national origin, age, sex, marital status, or public assistance status \cite{ftc-2021}.

The Biden Administration launched the National AI Research Resource Task Force \cite{aigov}, as directed by Congress in the National AI Initiative Act of 2020, to advance AI research and explore AI's implications \cite{bidenAItaskforce}. Following this, additional legislation such as the Algorithmic Fairness Act of 2020 \cite{ai_fairness_act} and Algorithmic Accountability Act of 2022 \cite{algorithmic_accountability_act_2022} emerged, and in 2021 alone, 17 states introduced AI regulations \cite{legislation_ai_states}. Notably, New York City enacted a law (effective in 2023) prohibiting the use of automated decision-making tools to screen job candidates without prior bias audits and transparency regarding their use \cite{nyc_ai_ban}. As of 2023, many U.S. states have introduced and enacted AI legislation \cite{ncslai}.

Legislation specific to fairness in the employment and hiring context is enforced by the U.S. Equal Employment Opportunity Commission (EEOC). These laws protect individuals against discrimination based on race, color, religion, national origin, sex, sexual orientation, or gender identity or expression, as stipulated in Title VII of the Civil Rights Act of 1964, amended by The Civil Rights Act of 1991. Additionally, the Pregnancy Discrimination Act amendment to Title VII protects women against discrimination due to pregnancy or childbirth. The Equal Pay Act of 1963 prohibits wage and benefits discrimination based on sex. The Age Discrimination in Employment Act of 1967 protects individuals 40 years of age or older against age discrimination. The Americans with Disabilities Act of 1990 (Title I) and Sections 501 and 505 of the Rehabilitation Act of 1973 safeguard qualified individuals with disabilities. The Genetic Information Nondiscrimination Act of 2008 prohibits discrimination based on genetic information \cite{us_eeoc}. In 2021, the EEOC launched an initiative to ensure that AI and tools used in hiring and employment comply with these laws \cite{eeoc_2021_ai}, and as of 2023, it has provided further guidelines for employers \cite{eeoc_ai_titlevii}.

Previously, for employee selection, the EEOC established a requirement to ensure the fair treatment of job candidates by preventing adverse impact due to hiring decisions \cite{barocas2016big, equal1979adoption}. According to EEOC guidelines, \textit{adverse impact} is determined by applying the four-fifths or eighty percent rule. This rule states that the selection rate for a protected group must be at least four-fifths (or 80\%) of the rate for the group with the highest selection rate \cite{equal1979adoption}. For instance, as illustrated in Table \ref{tbl:selection_sample}, if the selection rate for Black applicants is 50\% of the selection rate for White applicants, it falls below the 80\% threshold, indicating an adverse impact \cite{equal1979adoption}. From this and other legislation, two key definitions of discrimination have emerged in AI literature: \textit{disparate treatment}, which refers to intentionally discriminatory practices targeted at individuals based on their protected attributes; and \textit{disparate impact}, which encompasses practices that result in a disproportionately adverse effect on protected groups \cite{equal1979adoption, zafar2019fairness, barocas2019fairness}. 

\begin{table}[!h]
\begin{center}
\begin{tabular}{ p{1.5cm}p{1.5cm}p{2cm}p{2cm} } 
\toprule
   \textbf{Applicants} & \textbf{Hired}& \textbf{Selection Rate}& \textbf{Percent Hired}\\
   \midrule
80 White & 48 & 48/80 & 60\% \\
40 Black & 12 & 12/40 & 30\% \\\bottomrule
\end{tabular}
\end{center}

\caption{Example illustrating selection rate disparity according to the EEOC's four-fifths rule, demonstrating adverse impact.}\label{tbl:selection_sample}
\vspace*{-.2in}
\end{table}

However, the four-fifths rule and prior legislation are only the first step towards establishing guidelines to assess a fair hiring process using AI. For instance, the four-fifths rule has come under prior scrutiny with proposed legislation advocating for a smaller disparity in selection rates \cite{cohenmilstein_ai_bias}. There have also been prior cases wherein these disparities have been contested on grounds of business necessity (that is, the criteria are directly job-related), cost considerations (alternatives are prohibitively expensive to implement), and qualification-based arguments (disparities result from differences in applicant qualifications rather than organizational practices)~\cite{CascinoMaatman2015,hart2006disparate}.

Furthermore, although it might seem straightforward to meet these standards by simply removing protected attributes from the training data, many situations still arise where, despite these efforts—and sometimes because of them—unfair decisions are made. For instance, in natural language processing (NLP), a model might infer a candidate’s gender from their name, a phenomenon known as \textit{indirect discrimination} or discrimination through features implicit in the dataset \cite{zafar2019fairness}. Similarly, legislation has often lagged behind AI advancements, with relevant laws only emerging in recent years and urging practitioners to act proactively. Therefore, achieving fairness in decision-making systems requires a nuanced approach that addresses the underlying causes of bias in the AI. This is discussed in the following section.

\subsection{Causes of Bias}
\label{sec:causes-of-bias}
There are several common ways in which human bias can be transferred to an AI model \cite{barocas2016big, mujtaba2019ethical, hardtFairness}:

\begin{enumerate}

  \item \textit{Training data:} If the training data is biased, an AI system trained on it will learn and propagate that bias. This can occur through a skewed sample, where one group is disproportionately represented in achieving a particular outcome compared to another, or through tainted label outcomes (e.g., if the dataset was manually labeled by humans, any inherent human bias may be transferred to the labels).
  
  \item \textit{Label definitions:} The \textit{target label} in AI represents the outcome a model aims to predict. Selecting a holistic or vague definition of the target outcome can result in a larger disparate impact. For example, if a manager is building a classification model to predict a job candidate as a ``good'' hire, various definitions could be applied to determine a ``good'' hiring decision \cite{barocas2016big}. Factors such as employee motivation, \textit{person-job fit} (i.e., the alignment of an applicant to a job), predicted tenure, or \textit{person-environment fit} (i.e., the compatibility of an applicant with the organization’s culture and workplace) typically determine how well a candidate fits an organization and their expected performance once hired \cite{krishnakumar2019assessing, barrick1991big, barocas2016big}. However, selecting a label with a systematically disproportionate value for members of a protected class will bias the model’s predictions \cite{barocas2016big}. For instance, using predicted tenure as a target label, which may seem reasonable to an employer, might result in bias against women if historical data shows lower tenure rates for women compared to men \cite{barocas2016big}.
  
 \item \textit{Feature selection:} \textit{Features} are attributes that describe or contribute to an outcome that an AI model uses during prediction. While feature selection can improve model accuracy by reducing data complexity and learning time \cite{verma2018fairness}, it can also lead to bias if the statistical variation between individuals in protected groups is not well-represented for a particular feature \cite{barocas2016big}. Features may lack the necessary granularity for an AI model to learn accurate classification for protected groups \cite{barocas2016big}. For instance, academic credentials and the college from which a job candidate graduated are commonly used in hiring; however, if members of a protected group graduate at a disproportionately higher or lower rate from certain colleges, an AI model might incorrectly classify job candidates from the protected group, even if they are qualified for the job \cite{barocas2016big}.
 
 \item \textit{Proxies:} Even with protected attributes removed from the training data, group membership can often be inferred from other attributes, potentially leading to biased decisions. For instance, word embeddings used to represent text in resumes have been shown to contain biased representations of gender, as evidenced by the distance between word vectors \cite{bolukbasi2016man}. For example, although the term ``computer programmer'' is gender-neutral, its underlying vector representation may be closer to gender-specific terms such as ``man,'' causing it to serve as a proxy for gender \cite{bolukbasi2016man}. However, in hiring, there are situations where proxies are necessary to verify credentials or to establish a bona fide occupational qualification (BFOQ)—for example, when a religious institution requires employees to be of a specific faith \cite{BFOQ_Cornell}. In such cases, protected attributes must be retained for fairness auditing, and potential bias mitigation techniques should be applied, as further described in Sec. \ref{sec:methods}.
 
 \item \textit{Masking:} If model developers or employers hold prejudicial views, they might collect data samples or engineer features that intentionally bias the AI model \cite{barocas2016big}. This practice is known as \textit{masking}, where a data representation (e.g., using only coarse-grained model features that obscure fine-grained group details) is employed to justify AI model bias toward certain protected groups \cite{barocas2016big}.
 
\end{enumerate}

\section{Bias in AI-Based Recruitment}
\label{sec:bias}

\begin{figure*}[t!]
\centering 
\includegraphics[width=\textwidth,trim={.25cm 8.2cm .2cm 9.2cm},clip]{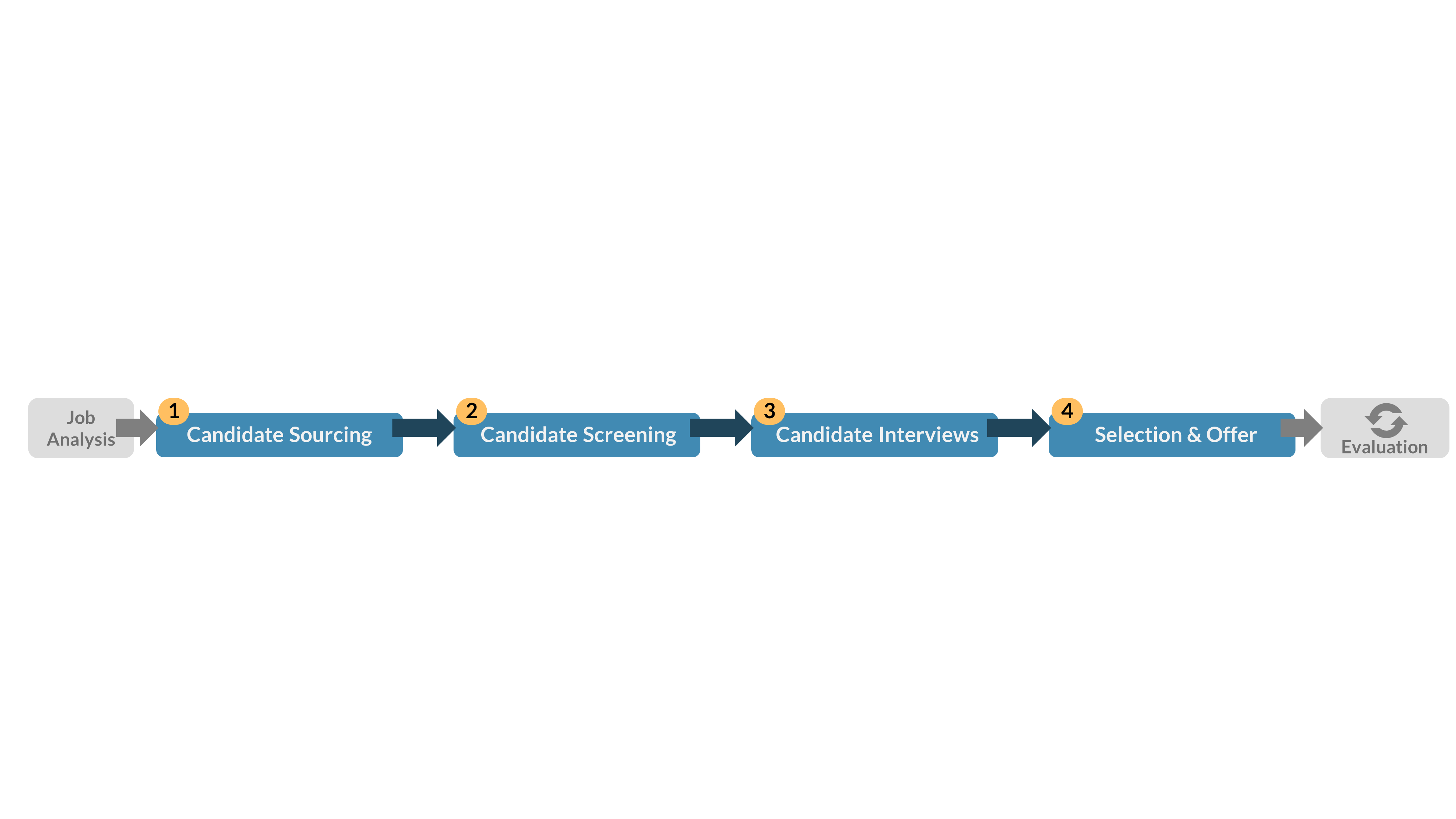}
\caption{Overview of steps in a typical AI-driven recruitment pipeline, including: (1) candidate sourcing (after job analysis), (2) candidate screening, (3) candidate interviews, and (4) selection and offer, with post-decision evaluation.}
\label{fig:pipeline2}
\end{figure*}

\begin{table*}[!t]
\begin{center}
\begin{tabular}{ p{1.5cm}|p{4cm}|p{1.5cm}|p{6cm}|p{3cm} } 
\toprule
   \textbf{Hiring Stage} & \textbf{Description} & \textbf{References} & \textbf{Summary of Key Challenges} & \textbf{Most Relevant Fairness Metrics}\\
   \midrule
\textbf{Candidate Sourcing} & Forming a job description through job analysis, writing a job ad, and outreach to potential candidates via job boards or networking sites. & \cite{hiretual,arya,textio_chatgpt,seekout_assist,talentgpt,brighthire,wilson2021building,tang2017gender,lin2021algorithmic,abououf2021machine,kurek2024zero,zheng2023generative,putka2023evaluating,bhola2020retrieving,micro1AI,linkedin,indeed,yang2017combining,almalis2015fodra,yagci2017ranker,miller_2015,carnevale2014understanding,ali2019discrimination,chen2018investigating,imana2021auditing,chen2019correcting,rus2022closing,li2023fairness,salinas2023unequal} & Potential biased language in job postings discourages diverse applicants, perpetuated by AI tools trained on historically biased data, further limiting applicant diversity. & Demographic parity, Counterfactual fairness, Multi-sided fairness \\
\arrayrulecolor{black!30}\midrule
\textbf{Candidate Screening} & Recruiters filter initial candidate pool by reviewing resumes and applications for job fit, typically without direct interaction. & \cite{hireVue,plum,seekout_assist,pena2020bias,pena2020faircvtest,deshpande2020mitigating,burke2021fair,wilson2024gender,seshadri2025does,abid2021persistent,bolukbasi2016man,kong2024gender,cowgill2018bias,meyer2018amazon,gan2024application,sunico2023resume,bevara2025resume2vec,kinge2022resume,ayishathahira2018combination,mhatre2023resume,singh2010prospect,wang2024jobfair,an2025measuring,an2025mutual} & AI models may prioritize quantifiable metrics over qualitative attributes, indirectly discriminate via proxies for protected attributes, and lack transparency in decision rationale. & Accuracy parity, Demographic parity, Counterfactual fairness, Individual fairness, Predictive rate parity \\
\arrayrulecolor{black!30}\midrule
\textbf{Candidate Interviews} & Interaction with candidates to evaluate detailed KSAOs, personality traits, and cultural fit via video, phone, or chat-based AI systems. & \cite{hireVue,brighthire,hirevuefacial2021,raghavan2020mitigating,wang2018adversarial,corbett2018measure,leong2019trust,jd_supra_2020,lee2019exploring,feine2019gender,junior2021person,yan2020mitigating,xue2023bias,booth2021integrating,lal2025exploring,boudjani2023ai,lee2024utilizing,han2021designing,hickman2024automated,sun2024facilitating,li2023ezinterviewer,hickman2022automated,moore2018whistle,mujtaba2024lost,zhang2024can,biswas2024hi} &
Biases embedded in large pre-trained models (LPTMs) from textual, audio, and visual data amplify implicit stereotypes, often resulting in accuracy disparities for non-native speakers or those with speech disabilities. Limited transparency and reliability concerns with automated personality assessments also increase regulatory scrutiny and fairness concerns. & Individual fairness, Predictive rate parity, Accuracy parity, Procedural fairness, Counterfactual fairness, Fairness in regression \\
\arrayrulecolor{black!30}\midrule
\textbf{Selection \& Offer} & Selecting final candidates, negotiating offers, and providing onboarding recommendations such as salary and benefits. & \cite{pactum,peng_investigations_nodate,oracleHCM,thida2023automated,geiger2025asking,sharma2024method,ji2025enhancing} & 
AI reliance on historical data perpetuates pay gaps, undervalues experience, penalizes resume gaps, and exacerbates information asymmetry during negotiation. Limited transparency in decision-making also negatively affects candidate perceptions of fairness. & Individual fairness, Predictive rate parity, Procedural fairness, Fairness in regression \\
\arrayrulecolor{black}\bottomrule
\end{tabular}
\end{center}

\caption{Summary of surveyed research, fairness challenges, and most relevant fairness metrics identified at each stage of the AI-driven recruitment pipeline.}\label{tbl:refernces_list}
\vspace*{-.2in}
\end{table*}

HR and I-O psychology literature typically divides the recruitment pipeline of an organization into four stages, as depicted in Figure~\ref{fig:pipeline2} \cite{holm2012recruitment, landy2016work, bogen2018help}. Each stage presents unique opportunities and challenges for bias introduction and mitigation. This section reviews current AI-driven practices at each stage, highlights potential sources of bias, and discusses existing efforts to achieve fairness throughout the recruitment process.

\subsection{Candidate Sourcing} 
The first stage of the recruitment process begins with a \textit{job analysis}, where an employer identifies a position within the organization and its required experience, knowledge, skills, abilities, and other pertinent characteristics (KSAOs). This stage also involves evaluating criteria for person-environment fit \cite{krishnakumar2019assessing, chapman2003use}, which includes assessing how well an individual’s values and traits align with the company’s work culture.

Following the job analysis, the employer moves to \textit{sourcing candidates}, which involves creating a pool of potential candidates by attracting job seekers to apply. This is accomplished by crafting job ads that reflect the identified KSAOs, conducting recruiter outreach, and leveraging employee referrals \cite{landy2016work, chapman2003use}.

In this context, AI-based methods have been developed for both job analysis and candidate sourcing stages. AI-based candidate sourcing platforms, such as HireEZ \cite{hiretual}, LinkedIn \cite{linkedin}, and Arya \cite{arya}, utilize AI to reach a broader and more diverse range of candidates. Other works, relying upon genetic algorithms \cite{abououf2021machine} or large pre-trained models (LPTM) \cite{kurek2024zero} have also been proposed. Additionally, LLMs are employed for various tasks, including crafting job ads \cite{textio_chatgpt}, conducting job analysis \cite{putka2023evaluating,bhola2020retrieving}, and sourcing candidates through prompting-based interfaces like TalentGPT \cite{seekout_assist,talentgpt}. They have also been used on the candidate side, by recommending jobs \cite{zheng2023generative}. Other AI-powered services, such as Textio, BrightHire, Micro1, and SeekOut \cite{wilson2021building,seekout_assist,brighthire,tang2017gender,micro1AI}, are also available to streamline the candidate sourcing process. The advantage to AI usage in this stage is the efficient processing of candidate data, though a 2021 study noted a lack of trust in data accuracy, control, and transparency \cite{lin2021algorithmic}.

Bias in AI-based candidate sourcing generally originates from two prominent sources. First, bias arises in the textual content of job advertisements, which often include gendered keywords, phrases, and language perpetuating age or racial stereotypes \cite{textio_chatgpt,salinas2023unequal,lambrecht2019algorithmic,kuhn2013gender,carpenter2015google,tang2017gender}. Such biased wording can deter certain demographic groups from applying, thereby limiting the diversity of the applicant pool. Services like Textio have sought to address this issue by employing AI-driven methods to suggest more neutral, inclusive language \cite{textio_chatgpt}. However, because many job advertisements remain untested for bias, this problem persists. Unchecked biases in job ads not only increase organizational time and resource expenditures in seeking diverse qualified candidates but also reduce the trust and \textit{self-efficacy}—the confidence of jobseekers in their ability to succeed in the recruitment process—particularly among members of underrepresented or marginalized groups \cite{woods2020personnel,wanberg2020job}.
Second, bias can stem from word embeddings in text-based models, stemming from inherent cultural, gender, racial, and political biases found in underlying training text data \cite{bolukbasi2016man}. Biased job ads can also influence AI-driven job-ad recommendations, leading to skewed ads being shown to users, as observed in previous studies on Facebook and Google job ad recommendations \cite{ali2019discrimination,miller_2015}.

Job recommendation systems, used by jobseekers to browse job ads, heavily rely on AI to match resumes with ads in their database, often leading to biased recommendations. Examples of such platforms include LinkedIn \cite{linkedin} and Indeed \cite{indeed}, where job postings are aggregated for jobseekers to search, while recruiters frequently search for \textit{passive candidates}—individuals who are not actively seeking a job but fit a job ad well \cite{landy2016work}. The recommendation algorithms on these sites typically use collaborative filtering \cite{yang2017combining}, content-based filtering \cite{almalis2015fodra,yang2017combining}, or multi-objective ranking approaches \cite{yagci2017ranker}. However, job ad content can lead to biased recommendations, as an estimated 60\%--70\% of job ads posted online disproportionately represent industries for high-skilled workers and STEM occupations \cite{carnevale2014understanding}. Moreover, ad delivery systems can be skewed along gender and racial lines, as evidenced by Facebook’s job recommendation engine in 2019 \cite{ali2019discrimination}. Additionally, gender inequality was observed on job boards such as Indeed, Monster, and CareerBuilder in 2018, with 12 of the 35 studied job titles showing significant group unfairness \cite{chen2018investigating}.

To address such biases, various approaches have been developed. For example, Imana \textit{et al.} distinguish between skew in ad delivery caused by protected attributes and skew due to differences in candidate qualifications \cite{imana2021auditing}. Similarly, Chen \textit{et al.} \cite{chen2019correcting} tackle recency bias in AI-driven information retrieval, where recent content disproportionately influences model training, leading to bias. Additionally, Rus \textit{et al.} \cite{rus2022closing} use a generative adversarial network to eliminate gender biases from word vector representations in job ad text and resumes. Meanwhile, Li \textit{et al.} model the job recommendation task as a resource allocation problem, promoting fairness by maintaining consistency across users from different protected groups \cite{li2023fairness}. Despite these efforts to mitigate cognitive biases in job and candidate recommendations, such biases may persist through the training data \cite{bogen2018help}. The issue of fairness in recommender systems extends beyond job recommendations and encompasses biases in e-commerce and content recommendation. It represents an active research area with ongoing efforts to address bias and enhance fairness \cite{jin2023survey}.

\vspace*{-.1in}

\subsection{Candidate Screening}

The next stage, \textit{candidate screening}, involves filtering a pool of candidate profiles (e.g., resumes and applications) based on their alignment with job requirements, or person-job fit \cite{krishnakumar2019assessing,landy2016work,chapman2003use}. To manage large volumes of applicants efficiently, organizations increasingly utilize AI-driven tools such as HireVue \cite{hireVue}, Plum \cite{plum}, and SeekOut \cite{seekout_assist}. These platforms offer automated resume ranking, candidate skill assessments, and preliminary evaluations of personality traits. Recent research has also explored advanced AI techniques, including traditional machine learning methods \cite{kinge2022resume,singh2010prospect}, deep learning models like convolutional neural networks \cite{mhatre2023resume}, recurrent neural networks \cite{ayishathahira2018combination}, and LLMs \cite{gan2024application,sunico2023resume,bevara2025resume2vec}.

However, similar to other areas of AI use, candidate screening models are susceptible to biases, as observed in high-profile cases such as Amazon’s AI-driven resume screening tool \cite{meyer2018amazon,cowgill2018bias}. Institutional biases inherent in historical hiring practices are frequently reflected in these models because their training data often lack diversity, perpetuating existing disparities \cite{bogen2018help}. Moreover, screening tools built upon NLP techniques are vulnerable to biases embedded within underlying training datasets, which can unintentionally reinforce harmful stereotypes or unequal representations of demographic groups \cite{bolukbasi2016man}. For instance, widely used word embedding models were found to contain significant gender biases due to their training on biased textual data from sources like Google News \cite{bolukbasi2016man}. Similarly, language models like GPT-3 have demonstrated biases against certain religious groups \cite{abid2021persistent}, and other prominent models, including GPT-4 and BERT, have faced scrutiny regarding biases related to gender and race \cite{gallegos2023bias}. Such biases in foundation models can directly lead to unfairness in resume screening and candidate selection processes, disadvantaging qualified candidates from underrepresented groups \cite{wilson2024gender,seshadri2025does,kong2024gender}. This bias not only undermines fairness but may also result in suboptimal hiring decisions, higher training costs, reduced productivity, and decreased employee retention and satisfaction, ultimately impacting organizational effectiveness and workplace diversity \cite{snell2006researching}.

Recognizing the importance of screening processes, many researchers have focused on developing methods to detect and mitigate biases. Recent work by Wang et al. \cite{wang2024jobfair} have studied gender hiring bias in approaches using LLMs for resume scoring, and find seven out of ten studied LLMs show significant biases against males in at least one industry studied \cite{wang2024jobfair}. Similar studies on GPT-4o, Google Gemini 1.5, and Claude 3.5 have been conducted by An et al. \cite{an2025measuring}, wherein they studied gender and racial bias in these models for evaluating resumes, finding lower resume scores were assigned for Black male candidates with similar work experience, education, and skills \cite{an2025measuring}. A similar study was conducted on LLM representations of occupations, and found perpetuation of gender stereotypes of jobs based on names; for instance, feminine first-name embeddings increased probabilities for female-dominated jobs \cite{an2025mutual}. Another approach by Pena \textit{et al.} \cite{pena2020bias,pena2020faircvtest} proposes a testbed for resume screening algorithms to identify bias in candidate scoring across five scenarios with different protected attributes. Similarly, Deshpande \textit{et al.} \cite{deshpande2020mitigating} tackle bias by reweighting term frequency--inverse document frequency values based on fairness metrics when matching resumes with job descriptions. Burke \textit{et al.} \cite{burke2021fair} introduce a mitigation method through candidate ranking. While these techniques represent an initial effort in bias mitigation, non-textual features within screening models, such as images (commonly found in online profiles), and key non-protected attributes like educational background may still contribute to perpetuating bias. This issue, where non-textual features such as visual and auditory cues contribute to bias, is particularly prevalent in the screening of candidate video or phone interviews.

\vspace*{-.05in}

\subsection{Interviewing Candidates}

The third stage in the recruitment process involves interviewing candidates to assess their KSAOs, work values, personality, and other traits in detail \cite{landy2016work,krishnakumar2019assessing}. Over the past decade, the recruitment process has increasingly relied on AI to evaluate these traits during interviews. Various AI modalities have been utilized, including chatbot systems for candidate interaction \cite{boudjani2023ai,lee2024utilizing,han2021designing}, automated video and phone interviews \cite{hireVue,hirevuefacial2021,hickman2024automated}, and online question-answer assessments \cite{hireVue,bogen2018help,raghavan2020mitigating,lal2025exploring}. LLMs have also been applied for mock interviews and scoring interview answers \cite{sun2024facilitating}. Similarly, language models have been used to prepare job interview questions based on resume content \cite{li2023ezinterviewer} and refine pre-defined questions \cite{shrmchatgpt}. Commercial platforms like HireVue \cite{hireVue}, Micro1 \cite{micro1AI}, and BrightHire \cite{brighthire} also heavily depend on AI to evaluate video interviews, scoring candidates based on personality traits and their fit for the job.

However, video- and image-driven AI interviewing systems have raised substantial fairness concerns, leading to both public and regulatory scrutiny, as highlighted in previous studies \cite{wang2018adversarial, corbett2018measure, leong2019trust}. For instance, HireVue discontinued its video-based candidate scoring systems in 2021, citing concerns over transparency and potential biases in AI decision-making \cite{hirevuefacial2021}. Illinois enacted the Artificial Intelligence Video Interview Act in 2020, becoming the first jurisdiction to regulate AI used in candidate interviews, requiring employers to disclose AI use, describe evaluation criteria, and obtain applicant consent before interviews \cite{jd_supra_2020}. Biases in these AI systems typically originate from training datasets containing implicit biases or unrepresentative samples, as observed in research into both video-interviewing and chatbot systems \cite{lee2019exploring, feine2019gender}. For example, the widely-used First Impressions dataset, employed in the 2017 ChaLearn challenge at CVPR, was later identified to contain labels biased against participants based on gender and race \cite{junior2021person, yan2020mitigating}. Such biases can be amplified through model predictions, potentially disadvantaging marginalized groups. Furthermore, automated personality trait assessments based on self-reported data have been found to lack reliability and validity, raising caution about their use in hiring decisions \cite{hickman2022automated}.

Similar biases have been identified in other AI-driven interview modalities, such as automatic speech recognition (ASR) systems used to transcribe candidate responses during interviews. For instance, recent research has revealed substantial accuracy disparities in ASR models when evaluating speech from non-native English speakers, people of color \cite{hickman2024automated}, individuals who stutter \cite{mujtaba2024lost}, and those with speech disabilities such as dysarthria \cite{moore2018whistle}. Text-based interview evaluation models relying on LLMs like GPT-4 have also demonstrated biased scoring of candidate responses across demographic groups, potentially disadvantaging certain candidates based on gender or race \cite{zhang2024can}. Furthermore, research on asynchronous video interviews has shown that altering the perceived demographics of interviewees can influence evaluations, undermining fairness perceptions of the interview process \cite{biswas2024hi}.

To mitigate these biases in AI-based candidate evaluation, researchers have proposed various techniques. For example, Yan et al. introduced adversarial learning approaches specifically aimed at reducing demographic bias in automated personality trait predictions during interviews \cite{yan2020mitigating}. Booth et al. explored bias measurement techniques within affective computing systems employed for automated analysis of video-based candidate responses \cite{booth2021integrating}. Additionally, significant research efforts have been directed toward improving fairness in chatbot interviewing systems by developing fairness-aware classifiers capable of detecting and mitigating harmful or biased textual interactions \cite{xue2023bias}.

\subsection{Selection and Evaluation} 

The final stage of the recruitment process involves selecting one or more job candidates, followed by negotiating an offer, which is then used to evaluate the organization's overall recruitment process \cite{landy2016work, krishnakumar2019assessing}. AI has been incorporated in several ways to assist not only in selection but also in the negotiation of job offers. For example, AI is used during job offer negotiations to provide salary and training recommendations, thereby aiding in the onboarding process of newly hired employees \cite{bogen2018help}. This is exemplified by software like Oracle Recruiting Cloud \cite{oracleHCM} which not only forecasts the probability of a job candidate accepting an offer but also allows employers to experiment with various salary structures, benefits packages, and other incentives to enhance the likelihood of acceptance \cite{bogen2018help}. Additionally, Pactum AI \cite{pactum} automates various negotiation processes, such as salary negotiations and contract discussions, leveraging a multitude of factors, including job market outlook. Related studies relying upon LLMs~\cite{thida2023automated,geiger2025asking} and neural networks~\cite{ji2025enhancing} have been published for forecasting job market demand and salary. Similarly, Sharma et al.~\cite{sharma2024method} proposed methods to forecast employee performance, work hours, and adjust salaries accordingly.

Although such AI seeks to create a fairer negotiation process by eliminating human biases, other forms of bias can still occur. For instance, a 2019 UpTurn report highlighted that Oracle and other AI-driven negotiation tools might reinforce asymmetric information between the job candidate and employer, potentially leading to an unfair process for the candidate \cite{bogen2018help}. Additionally, these tools may conflict with legislation prohibiting the use of salary history in hiring decisions \cite{bogen2018help}.

Job candidates who are not hired often face a lack of transparency regarding the AI-driven decision-making processes of employers. According to a survey conducted by The Talent Board, 53.5\% of candidates reported receiving no feedback after the screening and interview stages, while 69.7\% received no feedback after being turned down \cite{balanceCareers}. Although employers are not legally obligated to provide explanations \cite{balanceCareers}, the absence of feedback can negatively impact the perceived fairness and trustworthiness of an AI-based recruitment process. Transparency is crucial at every stage of the process, including communicating the job outcome, as it enables candidates to learn and improve for future interviews. Previous studies have demonstrated that providing explanations for organizational decisions can significantly enhance trust \cite{shaw2003justify, landy2016work}. Recently, many approaches, as further described in Section \ref{sec:methods}, have prioritized model interpretability and explainability to enhance transparency in AI-based decision-making, a principle that can be extended to AI-driven hiring decisions. The approach by Ji et al. \cite{ji2025enhancing} propose a neural network for explainability of salary predictions.

After finalizing hiring decisions, organizations evaluate their recruitment process based on three key criteria: validity, utility, and fairness. \textit{Validity} ensures that assessments and interviews accurately reflect job KSAOs and predict job performance \cite{landy2016work}. For instance, organizations often establish a \textit{criterion-referenced cut score} as a cut-off for employee selection, based on the desired level of performance for a new employee \cite{landy2016work}. Compliance with government guidelines, as described in Section \ref{sec:background:legislation}, is also essential to validity \cite{landy2016work}. Next, \textit{utility} assesses the cost-benefit ratio of staffing, which can be influenced by earlier recruitment stages such as employee outreach and job marketing. Finally, the \textit{fairness} criteria is evaluated using employee selection guidelines described in Section \ref{sec:background:fairness}. However, as AI becomes more integrated into each stage of the process, it is necessary to incorporate bias detection and mitigation methods where guidelines may not adequately address algorithmic bias. For example, the uniform guidelines on employee selection could be substantially affected by the sample size of different groups. Work by Peng \textit{et al.} \cite{peng_investigations_nodate} demonstrates how AI-human, or hybrid, decisions where AI assists humans, can still carry bias. Moreover, legislative definitions of fairness and employee selection guidelines may not be sufficiently detailed or up-to-date to address recent technological advancements.

\section{Bias Detection and Mitigation}\label{sec:detection}
Ensuring fairness in AI-driven hiring requires systematic methods for detecting, measuring, and mitigating bias at each stage of the recruitment pipeline. In this section, we review the existing work in AI fairness, focusing on methods for measuring bias, strategies for mitigating bias, and the current limitations and challenges in this field.

\subsection{Fairness Metrics}\label{sec:metrics}
Different fairness metrics have been proposed within AI fairness research, each capturing distinct dimensions of fairness. As discussed later in this subsection, fairness metrics often involve inherent trade-offs that require careful consideration. This subsection reviews core fairness metrics that are particularly relevant for recruitment and hiring contexts, drawing upon foundational literature in algorithmic fairness \cite{barocas2016big, hardtFairness, verma2018fairness}.

\begin{enumerate}

    \item \textit{Fairness through unawareness:} This metric provides a baseline definition of fairness, asserting that an algorithm is fair if it does not explicitly consider protected attributes ($A$) during training or decision-making \cite{kusner2017counterfactual}. Formally, a classifier $C: X \setminus A \to Y$ is trained on feature set $X$ excluding the protected attributes $A$, where $Y$ represents possible outcomes. Although intended to avoid direct discrimination by ensuring the model is ``unaware'' of protected characteristics, this approach has several limitations. Most notably, features excluded from $X$ can often be indirectly inferred through other correlated attributes, termed proxies. In a hiring scenario, even if gender is not directly included, an indicator such as attendance at a women-focused coding bootcamp might implicitly reveal the applicant's gender, thereby introducing unintended biases. Similarly, the prestige or location of educational institutions might indirectly encode socioeconomic status or ethnicity. 
    Therefore, achieving fairness through unawareness alone is insufficient, and expert knowledge remains crucial for accurately identifying and addressing such hidden biases across different occupational contexts.

    \item \textit{Demographic parity:} Also known as statistical parity, this metric stipulates that the acceptance rates for different demographic groups should be approximately equal. This requirement aligns closely with the concept of adverse impact as defined by the EEOC’s four-fifths rule, detailed in Section~\ref{sec:background:legislation}. Mathematically, let \( C \) be a classifier predicting a binary outcome \( Y \). For any two groups \( a \) and \( b \), the acceptance rates must be within a specified percentage \( p \) of each other, where \( \epsilon = \frac{p}{100} \) and \( \epsilon \in [0,1] \) \cite{hardtFairness}:
    \begin{equation}\label{eq:demographic_parity}
    \left| P_a\{C=1\} - P_b\{C=1\} \right| \leq \epsilon.
    \end{equation}

    \item \textit{Accuracy parity:} Also known as equality of opportunity, this metric ensures that the true positive rate (the proportion of actual positives correctly identified by the model) is equal across different demographic groups \cite{barocas2019fairness}. This form of fairness seeks to equalize the likelihood of a qualified candidate from any group being hired (i.e., \(C = 1\) when \(Y = 1\)). This concept is mathematically represented by the equation \cite{mujtaba2019ethical}:
    \begin{equation}\label{eq:accuracy_parity}
    P_a\{C=1 \mid Y=1\} = P_b\{C=1 \mid Y=1\}.
    \end{equation}
    This metric emphasizes fairness by ensuring qualified candidates from all groups have an equal probability of selection. However, it does not directly address potential disparities in false-positive or false-negative rates, which can still lead to unequal outcomes across groups.

    \item \textit{Predictive rate parity:} Also known as positive predictive parity and negative predictive parity, this metric ensures that the credentials of a candidate are consistent with the model's predictions across different groups \cite{mujtaba2019ethical}. Predictive rate parity is satisfied if both positive and negative outcomes are equally well predicted across groups. Positive predictive parity requires that the probability of a positive classification given a positive label is the same for all groups:
    \begin{equation}\label{eq:predictive_rate_pos}
    P_a\{C=1 \mid Y=1\} = P_b\{C=1 \mid Y=1\}.
    \end{equation}
    Negative predictive parity requires that the probability of a negative classification given a negative label is the same across groups:
    \begin{equation}\label{eq:predictive_rate_neg}
    P_a\{C=0 \mid Y=0\} = P_b\{C=0 \mid Y=0\}.
    \end{equation}
    These conditions help ensure that the model’s predictions consistently reflect actual candidate qualifications across different groups, preventing systematic biases that could unfairly advantage or disadvantage specific demographics.

    \item \textit{Counterfactual fairness:}     Counterfactual fairness requires that algorithmic decisions remain consistent even if a protected attribute \(A\) were hypothetically changed, thus ensuring robustness against discrimination based solely on protected characteristics. Defined by Kusner et al. \cite{kusner2017counterfactual}, a classifier \(C\) achieves counterfactual fairness if, for any individual with features \(X\) including a protected attribute \(A\), the outcome remains the same in both the actual and a counterfactual scenario where \(A\) is altered. The mathematical definition is:
    \begin{multline}\label{eq:counterfactual}
    P\{C_{A \leftarrow a}(X) = y  \mid  X, A=a\} \\ = P\{C_{A \leftarrow a'}(X) = y  \mid  X, A=a\},
    \end{multline}
    where \(a\) represents the actual value and \(a'\) represents a counterfactual value of \(A\), and \(y\) is the potential outcome. This ensures the prediction is independent of \(A\), except for its legitimate influences via other non-protected attributes.

    \item \textit{Individual fairness (versus group fairness):} Unlike group fairness metrics such as demographic parity, accuracy parity, and predictive rate parity which ensure fairness by equalizing statistical measures across groups, individual fairness focuses on fairness at the individual level. This metric asserts that similar individuals should receive similar outcomes. It addresses the key limitation of group fairness metrics, which ensure fairness across groups but may still permit unfairness within groups \cite{dwork2012fairness}. 
    Individual fairness requires that any two individuals who are comparable regarding relevant qualifications and characteristics for a job should receive similar outcomes from the model \cite{kim2019preference}.

    \item \textit{Preference-informed individual fairness:} 
    This fairness metric extends individual fairness by considering the diverse preferences of individuals, such as desired job location, industry, salary expectations, and work hours. Preference-informed individual fairness (PIIF) incorporates the concept of \textit{envy-freeness}, which means no individual would prefer another's outcome over their own under similar circumstances \cite{kim2019preference}. PIIF thus requires that outcomes not only meet individual fairness standards but also respect the personal preferences of individuals, ensuring decisions made by the model are both fair and aligned with individual job-related priorities.

    \item \textit{Multi-sided fairness:} 
    Modern recruitment platforms involve multiple stakeholders—employers posting job advertisements (producers), jobseekers (consumers), and the platform itself (the intermediary system). A multi-sided fairness approach considers fairness simultaneously from the perspective of all these groups \cite{patro2020fairrec}. Focusing exclusively on one stakeholder can unintentionally disadvantage others. For instance, a consumer-focused fairness approach (C-fairness) might mandate that job advertisements display similar salary ranges to all jobseekers, regardless of their protected group membership. Conversely, a producer-focused fairness approach (P-fairness) might require that advertisements from minority-owned businesses are shown to qualified jobseekers at comparable rates to those from other businesses \cite{burke2018balanced,burke2017multisided}. Multi-sided fairness is thus often formulated as a multi-objective optimization challenge, balancing accuracy and fairness across stakeholders, as seen in various online platforms such as e-commerce and ride-hailing services \cite{patro2020fairrec,suhr2019two,burke2018balanced,burke2017multisided}.

    \item \textit{Procedural fairness:} 
    Procedural fairness, introduced in Section~\ref{sec:background:fairness}, pertains to perceptions regarding the fairness of the processes used to arrive at decisions, such as candidate selection methods in recruitment. In the context of AI-based hiring, procedural fairness can be evaluated by assessing stakeholder perceptions about the legitimacy and appropriateness of features and criteria used by AI systems in making decisions. For instance, Grgic-Hlaca et al. \cite{grgic2018beyond} propose that procedural fairness in algorithmic decisions should reflect whether stakeholders view the use of specific features as justified and unbiased. However, perceptions of fairness can evolve as stakeholders become more aware of potential biases and unintended consequences associated with certain decision-making processes or features used by AI models.

    \item \textit{Fairness in regression:}
    Fairness metrics have traditionally focused on binary classification outcomes (e.g., hire or reject), yet regression models are prevalent in hiring contexts, such as predicting job performance scores from candidate interviews or resumes \cite{kaya2017multi}. In fairness analyses, the actual outcome (\(Y\)) differs from the model’s predicted outcome (\(R\)), with the prediction ideally closely approximating the true outcome. Here, the protected attribute (\(A\))—such as race, gender, or age—is considered when assessing potential biases in these predictions. According to Barocas et al. \cite{barocas2019fairness}, fairness in regression is guided by three principles:
    \begin{itemize}
    \item \textit{Independence}—ensuring the protected attribute \(A\) and the regression outcome \(R\) are statistically independent (\(R \perp A\)), indicating no influence of protected attributes on predictions.
    \item \textit{Separation}—permitting justified correlations between the protected attribute \(A\) and the target variable \(Y\), but ensuring predictions \(R\) remain independent of \(A\) given \(Y\) (\(R \perp A \mid Y\)).
    \item \textit{Sufficiency}—ensuring predictions fully represent relevant characteristics, such that outcomes
    \(Y\) are independent of protected attributes \(A\) given the predictions \(R\) (\(Y \perp A \mid R\)) \cite{barocas2016big}.
    \end{itemize}
    Moreover, regression fairness can be quantitatively evaluated using metrics like the Kullback-Leibler divergence, assessing differences in predictive distributions across demographic groups \cite{feng_has_nodate}.

\end{enumerate}

\textbf{Metric Selection:} Each of the above metrics addresses different aspects of fairness and is applicable in various scenarios within the recruitment process to ensure an equitable hiring process. Selecting the right fairness metrics for AI-driven hiring requires careful evaluation of both organizational objectives and the inherent trade-offs between different definitions. For instance, counterfactual fairness states an algorithm is fair if the hiring decision would remain unchanged for a candidate solely if their protected attribute were altered, while demographic parity stipulates the acceptance rates for different demographic groups should be approximately equal. These metrics may compete with one another and optimizing for one can compromise another aspect of fairness or accuracy. Moreover, fairness metrics developed in academic and technical contexts do not always align perfectly with legislative requirements. Laws such as Title VII of the Civil Rights Act establish broad non-discrimination baselines without prescribing specific numerical thresholds or computational formulas, leaving room for interpretation. Therefore, it is essential to align fairness metrics with both legal standards and operational goals. A multidisciplinary approach also involving legal experts and ethicists can help organizations navigate these differences and choose or develop measures that balance multiple fairness dimensions while complying with regulatory obligations. Experts in hiring and the specific job area are also needed to identify protected attributes and potential proxies that could contribute to bias. Once these protected attributes are determined, fairness metrics should be chosen to align with the hiring task. For example, during resume screening, metrics such as demographic parity and counterfactual fairness can be used to ensure selection rates are consistent across groups; however, for video interviews or later stages, a more fine-grained metric—such as accuracy parity—may be better suited to ensure candidates are adequately qualified. Additionally, ongoing bias evaluations should be conducted, as definitions of fairness and legislation evolve over time.

\subsection{Bias Mitigation Strategies}\label{sec:methods}

After detecting bias within an AI system, the next step involves implementing strategies for mitigation. This subsection outlines three primary categories of bias mitigation techniques as identified in existing literature \cite{hort2024bias}:

\begin{enumerate}
    \item \textit{Pre-processing:} This approach involves modifying the dataset before the model training process begins. Techniques in this category, such as reweighting and optimized pre-processing \cite{li2022achieving}, adjust features and labels in the dataset to align with fairness criteria before the model is trained and used for prediction \cite{hort2024bias}. These methods aim to remove or modify bias-inducing factors in the data, such as unrelated protected attributes, potential proxies of protected attributes, or imbalances in groups between training and test splits. Toolkits like AI Fairness 360 offer a range of pre-built functions for these pre-processing tasks \cite{bellamy2018ai}. Other recent techniques such as generation of synthetic data to balance representation in training datasets have also been proposed \cite{melzi2023synthetic}. However, while effective at data-level adjustments, pre-processing does not rectify biases inherent to the core modeling process, which may still require attention if biases persist after dataset adjustments.

    \item \textit{In-processing/optimization:} Unlike pre-processing, in-processing methods involve optimizing the model during training to meet specific fairness definitions through constraints or modifications in the model's design. This could, for instance, involve implementing accuracy parity by ensuring that a proportionate number of qualified candidates from different demographic groups are hired. In-processing techniques fall into two categories: explicit methods, which directly alter the training objectives, and implicit methods, which modify the representation of data within the model \cite{wan2021modeling}. A comprehensive discussion on these methods can be found in the survey by Wan et al. \cite{wan2021modeling}. The primary advantage of in-processing is its effectiveness in optimizing fairness metrics; however, this may sometimes come at the expense of reduced model accuracy \cite{kleinberg_inherent_2016}. Notably, in-processing presents challenges such as the necessity to alter the model architecture, which may not always be feasible, particularly in outsourced recruitment scenarios. Additionally, if the dataset's target labels are inherently biased due to systemic issues, these biases may continue to be propagated, despite the application of fairness constraints during model training.

    \item \textit{Post-processing/transparency:} Post-processing techniques focus on adjusting the outputs of an already-trained model to satisfy fairness criteria. For instance, this might involve modifying the classification thresholds used to make hiring decisions, or even modifying the predicted outcomes themselves \cite{hort2024bias}. The primary advantage of this approach is that it does not necessitate changes to the underlying model architecture. Additionally, these methods often incorporate elements of algorithmic transparency, such as providing explanations for decisions through counterfactual reasoning or analysis of feature importance. Revealing the decision-making process enables a thorough assessment of its fairness and allows for necessary adjustments if biases are identified. 
    Moreover, providing clear explanations helps job candidates understand potential discrepancies or deficiencies in their applications, which can be beneficial for their future job searches. Research has shown that transparent feedback can significantly improve candidates’ perceptions of fairness within the recruitment process \cite{gilliland1993perceived}.
\end{enumerate}

\textbf{Selecting a Bias Mitigation Strategy:} Upon discovering bias or unfairness in an AI system at any stage of the recruitment process, a bias mitigation strategy should be developed. The first step is to assess the underlying cause of the bias, which will guide the application of mitigation techniques. For instance, if an imbalance among groups in the dataset is contributing to bias, pre-processing techniques may be employed. However, it is important to consider the feasibility of different approaches for various researchers and practitioners. For example, pre-processing requires retraining the model, which incurs computational costs, while in-processing may not be feasible if the AI model is outsourced.

Another crucial aspect of bias mitigation is the continuous evaluation of bias even after a mitigation strategy has been applied. This ensures consistent fairness as hiring models evolve, practices change over time, and local legislation is followed. Numerous ML toolkits have been developed to integrate fairness algorithms at various stages of the ML pipeline. These toolkits—comprehensively reviewed in the survey by Lee et al. \cite{lee2021landscape}—support the implementation of pre-processing, in-processing, and post-processing methods. Their availability enables organizations to more easily embed fairness into their AI-driven hiring processes, thereby helping ensure that their recruitment practices meet established fairness standards. For a more detailed overview of bias mitigation strategies, we point readers to the survey by Hort et al. \cite{hort2024bias}.

\section{Limitations, Challenges, and Future Directions}
\label{sec:future}
This section outlines key limitations and challenges in current approaches to bias detection and mitigation in AI-driven recruitment, highlighting the need for continued research and development in this domain:

\begin{enumerate}
    \item \textit{Selecting metrics and auditing practices:} As organizations implement fairness practices, a pivotal challenge arises in selecting appropriate fairness metrics, which can vary significantly across different applications. In recruitment, for example, it is crucial to consider both candidate qualifications and the representation of protected groups. Additionally, the difficulty of meeting multiple fairness criteria simultaneously \cite{kleinberg_inherent_2016} suggests that relying on a single metric might ignore other significant disparities. Equally concerning is the absence of standardized benchmarks and comprehensive auditing guidelines in existing legislation, both essential for consistently assessing and ensuring fairness in AI recruitment tools \cite{pena2020faircvtest}.
    
    \item \textit{Job-specific requirements and long-term fairness:} The criteria for measuring fairness must adapt to the unique requirements of different jobs. For instance, although religion is generally considered a protected attribute, it may be legitimately relevant for leadership roles within religious organizations. This underscores the complexity of employing universal AI models across diverse job types, particularly when certain protected attributes can be legitimately job-relevant. Additionally, the need for fairness extends to job advertisements and their recommendations, which must align with the detailed requirements specified in the job descriptions. As the job market evolves, the models and auditing mechanisms must also be dynamic, adapting to long-term changes in the market. This dynamic aspect of fairness, focusing on long-term fairness impacts and adaptability, has been underexplored in the fairness literature, which has traditionally focused on static fairness assessments.
    
    \item \textit{Hiring decision transparency:} Many applicants who are screened out by AI-based systems never receive feedback, and even when feedback is provided, it often lacks sufficient detail or clarity regarding the reasons behind the decision.
    This lack of detailed feedback on unmet requirements or the features influencing the AI's decision undermines interactional justice and erodes trust. Enhancing the interpretability and explainability of black-box AI models is crucial for fostering transparency and fairness in AI-driven hiring. Despite emerging legislation aimed at improving transparency in AI-driven decisions, such initiatives are still limited and often lack robust enforcement mechanisms.
\end{enumerate}

These outlined challenges highlight the complex, evolving landscape of AI-driven hiring, underscoring the critical need for continued innovation and rigorous evaluation to advance fairness and transparency in recruitment practices. To address these challenges, we propose four potential future directions:
\begin{enumerate}
    \item \textit{Standards for AI Hiring:} Establishing rigorous standards or legislation for AI-driven hiring in government is essential to ensure that recruitment processes are fair. As organizations increasingly turn to AI to streamline candidate screening and decision-making, these systems must be held to robust ethical and legal benchmarks with clear guidelines outlining requirements for candidate rights to data privacy, using transparent AI-driven decisions with explanations, and implementing bias mitigation techniques. The EU's proposed AI Act \cite{eu_ai_act_2021} already categorizes AI usage in hiring as a high‐risk application, highlighting the need for rigorous regulatory standards. 
    Implementing similar legal frameworks in other jurisdictions is essential for establishing globally consistent standards for fair AI-driven recruitment.
    The Algorithmic Accountability Act of 2022, that requires companies to assess the impact of their automated decision-making systems, is a step forward, although is still in its proposal stages and does set forward explicit guidelines for nuanced bias measurement and mitigation in hiring \cite{algorithmic_accountability_act_2022}.

    \item \textit{Bias auditing large pre-trained models:} Large foundation models like ChatGPT, Whisper, and DALL-E are increasingly employed across various hiring tasks due to their robust performance in diverse scenarios. 
    These models utilize extensive training datasets comprising text, images, audio, and video data, each modality introducing distinct fairness challenges and biases.
    However, their widespread use raises significant ethical concerns due to the current lack of comprehensive regulations that ensure fairness throughout the model training and deployment phases. Key issues include preventing the leakage of sensitive data when generative models create content and promoting diversity in training datasets to minimize the perpetuation of stereotypes and harmful biases, thereby safeguarding equal treatment of all individuals \cite{hao2023safety}.

    \item \textit{Extend access to AI:} Extending access, or democratizing, AI tools to all researchers—even those with limited resources—is crucial for improving fairness in AI-driven hiring practices. When a diverse pool of researchers can independently audit, test, and refine AI hiring systems, it ensures that bias mitigation techniques are reflective of a wide array of perspectives \cite{bender_dangers_2021}. However, to train state-of-the-art large foundation models, there is a large and growing computational, financial, and environmental cost barrier to entry for many researchers \cite{bender_dangers_2021}. Open access, or open-sourcing of models, accompanied by clear ethical usage guidelines and transparency requirements, can foster broader community participation, enabling independent audits that identify and mitigate discriminatory practices embedded within proprietary AI systems.

    \item \textit{Including human feedback with AI:} Ensuring that human judgment remains a vital part of the hiring process is crucial to prevent the pitfalls of fully automated, AI-driven systems. By mandating that a certain percentage of hiring decisions involve human oversight, such as human resource experts, auditors, and hiring managers, organizations can mitigate potential biases in algorithmic models and offer nuanced decisions that models may miss. Properly trained human evaluators can provide contextual insights and ethical judgment that AI systems often struggle to capture fully, thereby facilitating a more holistic and fair assessment of candidates. This collaborative approach creates a continuous feedback loop, enabling AI systems and human evaluators to iteratively refine their practices and adapt to evolving standards of fairness and ethical hiring.
    
\end{enumerate}

\section{Conclusion}
\label{sec:conclusion}

In this paper, we systematically reviewed existing research on algorithmic fairness in AI-based recruitment systems, highlighting key complexities and challenges at the intersection of artificial intelligence, industrial-organizational psychology, and employment law. We began by clarifying essential concepts, including fairness, trust, and justice, grounding our discussion in interdisciplinary literature to establish a robust foundation for analyzing bias in AI hiring. We then examined the recruitment pipeline, discussing the role of AI at each stage, methods for detecting and mitigating bias, and the existing limitations and challenges highlighted in recent fairness literature. 

We identified significant areas in need of future research, including the development of AI models that incorporate fairness metrics tailored to job-specific requirements, the creation of fair job advertisements to engage a broader audience, and enhancing transparency in recruitment decisions. Additionally, pinpointing the origins of bias remains a pivotal challenge for the advancement of fair AI practices in recruitment.

The prevalence of bias in AI recruitment models poses considerable risks to organizations, job seekers, and the broader job market. Despite the enhanced efficiency provided by AI in screening vast numbers of applications, the propensity of these systems to perpetuate existing human biases is a significant concern. Addressing these issues requires robust enhancements in AI recruitment practices, incorporating thorough fairness auditing, transparent AI decision-making processes, and measures ensuring that job-specific requirements do not introduce further biases. Moreover, as the development of large foundation models like ChatGPT, Whisper, and DALL-E continues, the protection of applicant data privacy becomes increasingly imperative during model training and refinement phases.

Despite these challenges, the advancements in large foundation models offer unprecedented opportunities to transform recruitment processes worldwide. This potential drives the need for concerted efforts among organizations, researchers, and policymakers to ensure that the evolution of the recruitment landscape is both innovative and equitable, potentially leading to more dynamic workplaces and a more engaged workforce.

\ifCLASSOPTIONcompsoc

  \section*{Acknowledgments}
\else

  \section*{Acknowledgment}
\fi
This material is based upon work supported by the National Science Foundation under Grant No. 1936857.

\ifCLASSOPTIONcaptionsoff
  \newpage
\fi
\bibliographystyle{IEEEtran}
\bibliography{refs}

\end{document}